\documentclass[conference]{IEEEtran}
\IEEEoverridecommandlockouts
\usepackage{cite}
\usepackage{amsmath,amssymb,amsfonts}
\usepackage{algorithmic}
\usepackage{graphicx}
\usepackage{textcomp}
\usepackage{xcolor}
\usepackage{tikz}
\usepackage{authblk}

\ifCLASSOPTIONcompsoc
    \usepackage[caption=false, font=normalsize, labelfont=sf, textfont=sf]{subfig}
\else
\usepackage[caption=false, font=footnotesize]{subfig}
\fi

\def\BibTeX{{\rm B\kern-.05em{\sc i\kern-.025em b}\kern-.08em
    T\kern-.1667em\lower.7ex\hbox{E}\kern-.125emX}}
\newcommand{\orangeline}{\tikz[baseline]{\draw[orange,solid,line width = 1.0pt](0,0.8mm) -- (4mm,0.8mm)}}
\newcommand{\blueline}{\tikz[baseline]{\draw[blue,solid,line width = 1.0pt](0,0.8mm) -- (4mm,0.8mm)}}

\newcommand{\redline}{\tikz[baseline]{\draw[red,solid,line width = 1.0pt](0,0.8mm) -- (4mm,0.8mm)}}

\DeclareMathOperator{\T}{\mathsf{T}}

\makeatletter
\newcommand{\linebreakand}{%
  \end{@IEEEauthorhalign}
  \hfill\mbox{}\par
  \mbox{}\hfill\begin{@IEEEauthorhalign}
}
\makeatother

\begin{document}

\title{mmWave Simultaneous Localization and Mapping Using a Computationally Efficient EK-PHD Filter}

\author[${\dagger}$]{Ossi Kaltiokallio}
\author[$\star$]{Yu Ge}
\author[${\dagger}$]{Jukka Talvitie}
\author[$\star$]{Henk Wymeersch}
\author[${\dagger}$]{Mikko Valkama}

\affil[${\dagger}$]{Unit of Electrical Engineering, Tampere University, Tampere, Finland}
\affil[$\star$]{Department of Electrical Engineering, Chalmers University of Technology, Göteborg, Sweden}


\maketitle

\begin{abstract}
Future cellular networks that utilize millimeter wave  signals provide new opportunities in positioning and situational awareness. Large bandwidths combined with large antenna arrays provide unparalleled delay and angle resolution, allowing high accuracy localization but also building up a map of the environment. Even the most basic filter intended for simultaneous localization and mapping exhibits high computational overhead since the methods rely on sigma point or particle-based approximations. In this paper, a first order Taylor series based Gaussian approximation of the filtering distribution is used and it is demonstrated that the developed extended Kalman probability hypothesis density filter is computationally very efficient. In addition, the results imply that efficiency does not come with the expense of estimation accuracy since the method  nearly achieves the position error bound.

\end{abstract}


\section{Introduction}

Multipath propagation commonly degrades the performance of wireless position systems when no information is available regarding the location of the incidence point \cite{shen2010}. This is the case when the information captured in the waveform of the received signal is not sufficient for resolving the non-line-of-sight (NLOS) components in space and time. The fifth generation (5G) and beyond networks are expected to use signals in the millimeter wave (mmWave) band and employ massive multiple input multiple output (MIMO) antenna arrays at both the base station (BS) and user equipment (UE) \cite{andrews2014}. The large bandwidth in the mmWave band results in high temporal resolution \cite{patwari2005}, whereas the large antenna arrays allow utilizing narrow beams which enables accurate spatial resolution in the angular domain \cite{larsen2009}. The high temporal and spatial resolution of mmWave MIMO systems enable resolving the NLOS components and harnessing them for position and orientation estimation \cite{shahmansoori2018}.

Beyond position and orientation estimation, the location of the point of incidence of a NLOS path can be estimated, based on snapshot observations \cite{mendrzik2019} or on observations as the user moves \cite{witrisal2016high}. When the locations of the incidence points are inferred with sufficient accuracy, a map of the environment emerges, which in turn can help to position the user, even when the line-of-sight (LOS) signal is blocked \cite{talvitie2019}. The process of jointly localizing a UE and creating a map is known as 
simultaneous localization and mapping (SLAM), often referred to as channel-SLAM, radio-SLAM, or multipath-aided SLAM in the ultra-wideband (UWB) and  mmWave context \cite{GentnerTWC2016,LeitingerTWC2019,wymeersch2018,kim2018}. Radio-SLAM has several challenges that must be addressed \cite{kim2020c}: i) the tracked incidence point can be misdetected due to limitations in the receiver and channel estimation routine; ii) clutter measurements can generate false alarms; and iii) measurement ambiguities can lead to situations where a wrong measurement is associated to the wrong incidence point.
Several methods have been proposed for radio-SLAM including geometry-based algorithms \cite{yassin2018,palacios2018}, belief propagation on factor graphs \cite{LeitingerTWC2019,wymeersch2018,kim2018}, and random finite sets (RFSs) \cite{kim2020a,kim2020b,kim2020c}. Formulating the problem using RFSs is an attractive option since the method can inherently deal with radio-SLAM challenges, including an unknown number of targets, clutter measurements, misdetections and unknown data association. Among RFS radio-SLAM methods,
a Rao-Blackwellized particle-based probability hypothesis density (RBP-PHD) filter was recently developed, showing simultaneous localization and mapping of a propagation environment comprising of reflecting surfaces and small scattering point objects \cite{kim2020c}. Since the computational complexity scales exponentially with the state dimension, the computational burden of the RBP-PHD is significant. Low-complexity alternatives were proposed in \cite{kim2020a,kim2020b} which utilize a cubature Kalman PHD (CK-PHD) filter for radio-SLAM. As a downside, the CK-PHD is inferior compared to the RBP-PHD in terms of tracking accuracy \cite{kim2020a}. 

In this paper, we propose an extended Kalman probability hypothesis density (EK-PHD) filter for radio-SLAM. We adopt the CK-PHD approach presented in \cite{kim2020a,kim2020b} and utilize a first order Taylor series based Gaussian approximation of the filtering distribution instead of relying on sigma point \cite{kim2020a,kim2020b} or particle \cite{kim2020c} based approximations. The resulting algorithm is very efficient and it is over 100 times faster than the CKF implementation proposed in \cite{kim2020a,kim2020b}. In addition, we demonstrate that efficiency does not come with the expense of estimation accuracy. It is shown that the proposed solution nearly achieves the position error bound, as long as a sufficient number of measurements are available for estimating the reflecting surfaces and scattering points.

\section{Models}

\begin{figure}[tb]
  \centering
  \centerline{\includegraphics[width=8.5cm]{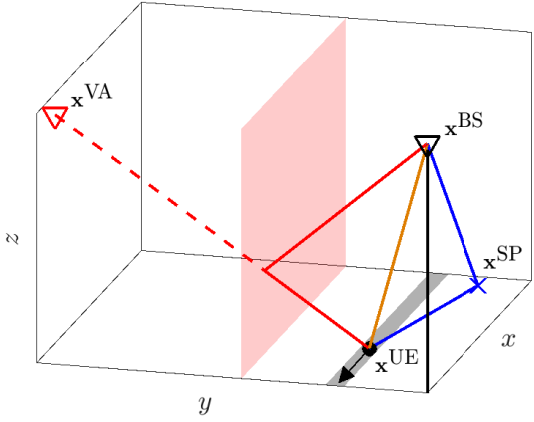}}
\caption{
The UE utilizes the LOS (\protect\orangeline), reflected (\protect\redline) and scattered (\protect\blueline) signals to jointly localize itself and build up a map of the environment. The mapped environment constitutes of landmarks. Reflection landmarks are characterized by virtual anchors (VA) and scattering landmarks are characterized by scattering points (SP).}
\label{fig:system_overview}
\end{figure}

The objective of this paper is to track the vehicle state $\mathbf{x}_k^{\text{UE}}$ and simultaneously map the surrounding environment by estimating location of the static landmarks $\mathbf{x}^{i}$. Since the number of landmarks, $n_k$, and measurement elements, $m_k$, are time varying, they are represented using random finite sets (RFSs). At time $k$, the map and measurements are $\mathcal{X}_k = \lbrace \mathbf{x}^{1}, \; \ldots, \; \mathbf{x}^{n_k} \rbrace$ and $\mathcal{Z}_k = \lbrace \mathbf{z}_k^{1}, \; \mathbf{z}_k^{2}, \; \ldots, \; \mathbf{z}_k^{m_k} \rbrace$ in respective order.

\subsection{State Model}

The considered scenario is illustrated in Fig.~\ref{fig:system_overview} in which one BS periodically sends a mmWave positioning reference signal (PRS). 
The PRS is received by a UE that is mounted on a vehicle. The location of the BS is known, $\mathbf{x}^{\text{BS}} \in \mathbb{R}^3$, whereas the landmark locations, $\mathbf{x}^{i} \in \mathbb{R}^3$, are unknown.  

The vehicle state comprises the vehicle position in the horizontal plane with respect to a fixed Cartesian coordinate frame, denoted by $x$ and $y$ [m]; $\alpha$ denotes the heading angle [rad]; and $B$ the unknown clock bias. As in related works \cite{kim2018,kim2020a,kim2020b,kim2020c}, it is assumed that elevation of the vehicle is known; and that magnitude of the velocity vector $v$ [m/s] and turn rate $\omega = \tfrac{d\alpha}{dt}$ [rad/s] are constant and provided by the electronic control module of the vehicle. The vehicle dynamics can be represented by a coordinated turn model \cite{roth2014} 
\begin{equation}\label{eq:dynamic_model}
     \underset{\mathbf{x}_{k+1}^{\text{UE}}}{\underbrace{\begin{bmatrix} x_{k+1} \\ y_{k+1} \\ \alpha_{k+1} \\ B_{k+1} \end{bmatrix}}} = \underset{\mathbf{f}\left(\mathbf{x}_{k}^{\text{UE}} \right)}{\underbrace{\begin{bmatrix} x_k + \tfrac{2v}{\omega}  \sin \! \left(\tfrac{\omega T}{2}\right)  \cos \! \left(\alpha_k + \tfrac{\omega T}{2}\right) \\ y_k + \tfrac{2v}{\omega}\sin \! \left(\tfrac{\omega T}{2}\right) \sin \! \left(\alpha_k + \tfrac{\omega T}{2}\right) \\ \alpha_k + \omega T \\ B_k \end{bmatrix}}} + \boldsymbol{\eta}_k,
\end{equation}
where $T$ is the sampling interval and  $\boldsymbol{\eta}_k \sim \mathcal{N}\left(\mathbf{0},\mathbf{Q}^\text{UE}\right)$ is the zero-mean process noise with covariance $\mathbf{Q}^\text{UE}$.

\subsection{Measurement Model}

At epoch $k$, the received signal at the UE can be expressed as \cite{heath2016}
\begin{multline}\label{eq:received_signal}
    \mathbf{y}_k\left( t \right) = \mathbf{W}_k^* \sum_{i=1}^{n_k} h_{k}^i \mathbf{a}_{\text{UE}}(\boldsymbol{\theta}_{k}^{R,i}) \mathbf{a}_{\text{BS}}^*(\boldsymbol{\theta}_{k}^{T,i}) \mathbf{B}_k  \mathbf{s}_k\left( t - \tau_{k}^i \right) \\ +  \mathbf{W}_k^* \tilde{\mathbf{r}}\left( t \right),
\end{multline}
where $i$ is the path index and $n_k$ the total number of resolvable paths, $\mathbf{W}_k$ and $\mathbf{B}_k$ are combiner and precoder matrices in respective order, $\mathbf{s}_k$ the PRS, $\tilde{\mathbf{r}}_k$ the noise, $h_{k}^i$ is a complex channel gain, $\mathbf{a}_{\text{UE}}$ and $\mathbf{a}_{\text{BS}}$ are antenna response vectors at the UE and BS sides respectively, and $\tau_{k}^i$,  $\boldsymbol{\theta}_{k}^{R,i}$ and $\boldsymbol{\theta}_{k}^{T,i}$ denote the time of arrival (TOA), direction of arrival (DOA) and direction of departure (DOD) in respective order. Both DOA and DOD have azimuth and elevation components, for example, $\boldsymbol{\theta}_{k}^{R,i} = {\begin{bmatrix} \theta_{\text{azimuth}}^{\text{DOA},i} & \theta_{\text{elevation}}^{\text{DOA},i} \end{bmatrix}}^{\T}$.

There exists several methods for estimating the TOA, DOA and DOD information from \eqref{eq:received_signal} such as sparse recovery \cite{shahmansoori2018}, compressed sensing \cite{talvitie2019} and subspace methods \cite{roemer2014}. However, channel estimation is outside the scope of this paper and it is assumed that the UE utilizes a channel estimation routine which directly outputs 
\begin{equation}
    \mathbf{z}_k^j = 
    \underset{\mathbf{g}\left(\mathbf{x}_{k}^{\text{UE}},\mathbf{x}^i \right)}{\underbrace{\begin{bmatrix}
    \tau_k^j & \theta_{\text{azimuth}}^{\text{DOA},j} & \theta_{\text{elevation}}^{\text{DOA},j} &
    \theta_{\text{azimuth}}^{\text{DOD},j} & \theta_{\text{elevation}}^{\text{DOD},j}
    \end{bmatrix}}}^{\T}
     + \mathbf{r}_k^j,
\end{equation}
where $\mathbf{r}_k^j \sim \mathcal{N}(\mathbf{0}, \boldsymbol{\Sigma}_k^j)$ is the measurement noise and $i$ is the unknown landmark index. It is important to note that the geometric model $\mathbf{g}(\mathbf{x}_{k}^{\text{UE}},\mathbf{x}^i )$ is a function of the UE state, as well as the landmark state. 
For LOS the signal source is the BS, whereas for NLOS the signal can either originate from specular reflection or scattering. The readers are referred to \cite{kim2020c} and the references therein for further details on $\mathbf{g}(\mathbf{x}_{k}^{\text{UE}} ,\mathbf{x}^i)$ and the different landmark types. See also Fig.~\ref{fig:system_overview}. 

Due to limitations in the receiver and channel estimation routine, it is possible that $\mathbf{z}_k^j \in \mathcal{Z}_k$ corresponds to clutter or $\mathbf{x}^i \in \mathcal{X}_k$ is not detected. In this paper, clutter is modeled with a Poisson point process (PPP) which is parametrized using intensity function $\lambda_c(\mathbf{z}) \geq 0$. In addition, adaptive detection probability, $P^\text{D}_k(\mathbf{x}_{k}^{\text{UE}},\mathbf{x}^i) \in [0, \; 1]$, and survival probability,  $P^\text{S}_k(\mathbf{x}_{k}^{\text{UE}},\mathbf{x}^i) \in [0, \; 1]$, are utilized to avoid problems with misdetected landmarks outside the field-of-view (FOV) \cite{wymeersch2020adaptive}. As in \cite{kim2020c}, it is assumed that $\mathbf{g}(\cdot)$, $\boldsymbol{\Sigma}_k^j$, $P^\text{D}_k(\cdot)$, $P^\text{S}_k(\cdot)$ and $\lambda_c(\cdot)$ are known to the vehicle.

\subsection{Performance Bounds} \label{sec:performance_bounds}

The Cram{\'e}r-Rao bound (CRB) for a time-varying system, usually referred to as posterior Cram{\'e}r-Rao bound (PCRB) \cite{tichavsky1995}, states that the mean squared error (MSE) of an estimator is always larger than $\mathbf{J}^{-1}$ in the positive semi-definite sense:
\begin{equation}\label{eq:cramer_rao_bound}
\mathbb{E}\lbrace \left(\hat{\mathbf{x}}(\mathbf{z}) - \mathbf{x}\right)\left(\hat{\mathbf{x}}(\mathbf{z}) - \mathbf{x}\right)^T \rbrace \geq \mathbf{J}^{-1},
\end{equation}
where $\mathbf{J}$ is the Fisher information matrix (FIM), $\hat{\mathbf{x}}(\mathbf{z})$ denotes an estimator of $\mathbf{x}$ which is a function of measurements $\mathbf{z}$. The FIM can be decomposed into two additive parts \cite{tichavsky1998}
\begin{equation}\label{eq:fim_partitioned}
    \mathbf{J} = \mathbf{J}_{\text{data}} + \mathbf{J}_{\text{prior}},
\end{equation}
where $\mathbf{J}_{\text{data}} = \mathbb{E}\left[-\Delta_\mathbf{x}^\mathbf{x} \log p(\mathbf{x}\vert\mathbf{z})\right]$ represents the information obtained from the data, $\mathbf{J}_{\text{prior}} = \mathbb{E}\left[-\Delta_\mathbf{x}^\mathbf{x} \log p(\mathbf{x})\right]$ represents the prior information and $\Delta_\mathbf{x}^\mathbf{x}$ is the second-order partial derivative. 

For the considered problem, the complete state of the system is $\breve{\mathbf{x}}_k = [(\mathbf{x}_k^\text{UE})^{\T} \; (\mathbf{x}^{1})^{\T} \; \ldots \; (\mathbf{x}^{n_k})^{\T}]$, the Jacobian matrices are given by
\begin{align}
    \lbrace \breve{\mathbf{G}}_{{\mathbf{x}}} \rbrace_{j,j'} &= \partial g_j\left( \breve{\mathbf{x}}  \right)/\partial \breve{x}_{j'}, \label{eq:G_Jacobian_UE} \\ 
    \lbrace \breve{\mathbf{F}}_{{\mathbf{x}}} \rbrace_{j,j'} &= \partial f_j \left( \breve{\mathbf{x}} \right) / \partial \breve{x}{j'}, \label{eq:F_Jacobian_UE}
\end{align}
and the process and measurement noise covariances are $\breve{\mathbf{Q}}_k = \text{blkdiag}(\mathbf{Q}^\text{UE},\mathbf{I}_{3 \times ({n_k-1})})$ and $\breve{\boldsymbol{\Sigma}}_k = \mathbf{I}_{n_k} \otimes \boldsymbol{\Sigma}$ in respective order and $\otimes$ denotes the Kronecker product. If the data association is known, the FIM can be recursively updated using 
\begin{equation}\label{eq:fim_recursive}
\mathbf{J}_{k} = 
   \underset{\mathbf{J}_{\text{data}}}{\underbrace{\breve{\mathbf{G}}_{\mathbf{x}}^{\T} \breve{\boldsymbol{\Sigma}}_k^{-1}\breve{\mathbf{G}}_{\mathbf{x}}}} +
   \underset{\mathbf{J}_{\text{prior}}}{\underbrace{\left(\breve{\mathbf{Q}}_k + \breve{\mathbf{F}}_{\mathbf{x}} \mathbf{J}_{k-1}^{-1} \breve{\mathbf{F}}_{\mathbf{x}}^{\T} \right)^{-1}}}.
\end{equation} 
From \eqref{eq:fim_recursive}, the position error bound (PEB) of the UE can be computed as $\text{PEB}_k = \sqrt{\lbrace \mathbf{J}_{k} \rbrace_{1,1}^{-1} + \lbrace \mathbf{J}_{k} \rbrace_{2,2}^{-1}}$ and the landmark error bound (LEB) of the $i$th landmark can be computed as $\text{LEB}_k^i = \sqrt{\sum_{j = 3i+2}^{3i + 4} \lbrace \mathbf{J}_{k} \rbrace_{j,j}^{-1}}$.  The computation of the Jacobians are straightforward but tedious, so they are omitted here for brevity.

\section{Estimators}

This work aims to compute the marginal posteriors of the vehicle and landmarks, which can be achieved by executing belief propagation on a factor graph representation of the joint posterior density \cite{wymeersch2018}. The joint posterior density, $p\left(\mathbf{x}_{0:K}^{\text{UE}},\mathcal{X}_{0:K} \vert \mathcal{Z}_{1:K} \right)$, can be factored as 
\begin{multline}\label{eq:joint_posterior_density}
   p\left(\mathbf{x}_{0:K}^{\text{UE}},\mathcal{X}_{0:K} \vert \mathcal{Z}_{1:K} \right) \propto 
   p\left(\mathbf{x}_{0}^{\text{UE}}\right) p\left(\mathcal{X}_{0}\right) \times \\ \prod_{k=1}^K  p\left(\mathbf{x}_{k}^{\text{UE}} \vert \mathbf{x}_{k-1}^{\text{UE}} \right)
   p\left(\mathcal{X}_{k} \vert \mathcal{X}_{k-1}, \mathbf{x}_{k}^{\text{UE}} \right)
   p\left(\mathcal{Z}_{k} \vert \mathbf{x}_{k}^{\text{UE}},\mathcal{X}_{k}  \right),
\end{multline}
where $p\left(\mathbf{x}_{0}^{\text{UE}}\right)$ and  $p\left(\mathcal{X}_{0}\right)$ are the priors, $p\left(\mathbf{x}_{k}^{\text{UE}} \vert \mathbf{x}_{k-1}^{\text{UE}} \right)$ and $p\left(\mathcal{X}_{k} \vert \mathcal{X}_{k-1}, \mathbf{x}_{k}^{\text{UE}} \right)$ the state transition densities and $p\left(\mathcal{Z}_{k} \vert \mathbf{x}_{k}^{\text{UE}},\mathcal{X}_{k}  \right)$ is the likelihood.   We adopt a Gaussian filtering approach for computing the marginal posteriors recursively \cite{kim2020a,kim2020b} and utilize a first order Taylor series based Gaussian approximation of the filtering distribution resulting in an algorithm that is computationally very efficient. In the following subsections, initialization, prediction and update steps of the developed estimator are introduced.

\subsection{Initialization}

The multiobject probability density function of an RFS, $p(\mathcal{X}_k \vert \mathcal{Z}_{1:k})$, is approximated using a PHD, $D_{k}(\mathbf{x})$, which propagates the first-order statistical moment of the RFS. The PHD, commonly referred to as intensity of the RFS, is parameterized using a Gaussian mixture (GM)
\begin{equation}
    D_{k}(\mathbf{x}) = \sum_{i=1}^{n_{k}} w_{k}^i \mathcal{N}\left(\mathbf{x} ; \boldsymbol{\mu}_{k}^i, \mathbf{C}_{k}^i \right),
\end{equation}
where $n_{k}$ is the number of targets at time $k$ and $w_{k}^i$, $\boldsymbol{\mu}_{k}^i$ and $\mathbf{C}_{k}^i$ are the weight, mean and covariance of the $i$-th GM component in respective order. The UE state is also approximated with a Gaussian density
\begin{equation}
\mathbf{x}_k^{\text{UE}} = \mathcal{N}\left(\mathbf{x}_k^{\text{UE}} ; \mathbf{m}_{k}, \mathbf{P}_{k} \right),
\end{equation}
where $\mathbf{m}_{k}$ and $\mathbf{P}_{k}$ are the mean and covariance of the estimate in respective order. 

The PHD is initialized with the BS position, $D_{0}(\mathbf{x})=\delta (\mathbf{x}-\mathbf{x}^{\text{BS}})$, and the UE state as $\mathbf{x}_0^{\text{UE}} = \mathcal{N}\left(\mathbf{x}_0^{\text{UE}} ; \mathbf{m}_{0}, \mathbf{P}_{0} \right)$. It is to be noted that during the first filter recursion, the prediction step is not computed so that the update step precedes the prediction step in the actual implementation. However, due to notational convenience, the prediction step is introduced first.

\subsection{Prediction}

\subsubsection{Vehicle Prediction}

Since time evolution of the vehicle does not depend on the landmarks, the predicted filtering distribution of the vehicle can be estimated independently. In this paper, a first order Taylor series based Gaussian approximation is used for which the series expansion can be formed as $\mathbf{f}(\mathbf{x}^{\text{UE}}) = \mathbf{f}(\mathbf{m} + \delta \mathbf{x}^{\text{UE}}) \approx \mathbf{f}(\mathbf{m}) +  \mathbf{F_x}(\mathbf{m})\delta \mathbf{x}^{\text{UE}}$, where $\mathbf{F_x}(\mathbf{m})$ is the Jacobian of $\mathbf{f}(\cdot)$ with elements $\lbrace \mathbf{F_x}(\mathbf{m}) \rbrace_{i,j} = \partial f_i (\mathbf{x}^{\text{UE}})/\partial {x}_j^{\text{UE}} \vert_{\mathbf{x}^{\text{UE}} = \mathbf{m}}$. Given the model, Jacobian and state estimate at the previous time step, $\mathcal{N}\left(\mathbf{x}^{\text{UE}} ; \mathbf{m}_{k-1}, \mathbf{P}_{k-1} \right)$, the first order extended Kalman filter \cite[Ch. 5.2]{sarkka_2013} can be used to predict the UE state at the current time step 
\begin{align}
    \mathbf{m}_{k \vert k-1} &= \mathbf{f}(\mathbf{m}_{k-1}), \\
    \mathbf{P}_{k \vert k-1} &= \mathbf{F_x}(\mathbf{m}_{k-1}) \mathbf{P}_{k-1} \mathbf{F}_\mathbf{x}^{\T}(\mathbf{m}_{k-1}) + \mathbf{Q}^\text{UE}.
\end{align}

\subsubsection{Map Prediction}\label{sec:map_prediction}

Let us assume that the map has a PHD, $D_{k-1}(\mathbf{x}) = \sum_{i=1}^{n_{k - 1}} w_{k-1}^i \mathcal{N}\left(\mathbf{x} ; \boldsymbol{\mu}_{k-1}^i, \mathbf{C}_{k-1}^i \right)$, at the previous time step. Then it follows that the predicted map is also a PPP with PHD \cite{vo2006}
\begin{equation}
D_{k \vert k-1}(\mathbf{x}) = \sum_{i=1}^{n_{k \vert k - 1}} w_{k \vert k-1}^i \mathcal{N}\left(\mathbf{x} ; \boldsymbol{\mu}_{k \vert k-1}^i, \mathbf{C}_{k \vert k-1}^i \right),
\end{equation}
which can be factorized as follows:
\begin{equation}
    D_{k \vert k-1}(\mathbf{x}) = D_{k \vert k-1}^\text{S}(\mathbf{x}) + D_{k-1}^\text{B}(\mathbf{x}),
\end{equation}
where $D_{k \vert k-1}^\text{S}(\mathbf{x})$ is the PHD of surviving objects and $D_{k-1}^\text{B}(\mathbf{x})$ is the birth process which captures all the appearing objects. The number of GM components in the predicted PHD is $n_{k \vert k-1} = n^\text{S}_{k \vert k - 1} + n^\text{B}_{k \vert k - 1}$, where $n^\text{S}_{k \vert k - 1}$ is the number of survived components and $n^\text{B}_{k \vert k - 1}$ the number of birth components.  
With approximate non-linear filters singularity of the static landmark's dynamic model can cause problems and the filter to diverge \cite[Ch. 12.3]{sarkka_2013}. To circumvent this problem, a small artificial noise is added and PHD of the surviving objects is given by
\begin{equation}
    D_{k \vert k-1}^\text{S}(\mathbf{x}) = \sum_{i=1}^{n_{k \vert k-1}^\text{S}} w_{k \vert k-1}^{\text{S},i} \mathcal{N}\left(\mathbf{x} ; \boldsymbol{\mu}_{k \vert k-1}^{\text{S},i} , \mathbf{C}_{k \vert k-1}^{\text{S},i}\right),
\end{equation}
where parameters of the GM are: 
$n_{k \vert k-1}^\text{S} = n_{k-1}$ number of surviving GM components;
$w_{k \vert k-1}^{\text{S},i} = P_{k-1}^\text{S}(\mathbf{x}) w_{k-1}^{i}$ weight of the component and $P_{k-1}^\text{S}(\mathbf{x})$ is an adaptive survival probability;
$\boldsymbol{\mu}_{k \vert k-1}^{\text{S},i} = \boldsymbol{\mu}_{k-1}^{i}$ and;
$\mathbf{C}_{k \vert k-1}^{\text{S},i} = \mathbf{C}_{k-1}^i + \tilde{\mathbf{Q}}^\text{MAP}$
where $\tilde{\mathbf{Q}}^\text{MAP}$ is the artificial process noise covariance. 

The birth PHD, $D^\text{B}_{k-1}(\mathbf{x})$,  is also represented as a GM
\begin{equation}
D^\text{B}_{k-1}(\mathbf{x}) =  \sum_{i=1}^{n_{k-1}^\text{B}} w_{k-1}^{\text{B},i} \mathcal{N}\left(\mathbf{x} ; \boldsymbol{\mu}_{k-1}^{\text{B},i} , \mathbf{C}_{k-1}^{\text{B},i} \right),
\end{equation}
where $n_{k-1}^\text{B}$ is the number of birth components and $w_{k-1}^{\text{B},i}$ the constant birth weight, $\boldsymbol{\mu}_{k-1}^{\text{B},i}$ the mean and $\mathbf{C}_{k-1}^{\text{B},i}$ the covariance. For each measurement, $\mathbf{z}_{k-1}^j$, that is not associated to an existing map object during the previous update step, a birth is generated for each possible landmark type, since the source of the measurement is unknown. Mean of the birth component is estimated using
\begin{equation}
    \boldsymbol{\mu}_{k-1}^{\text{B},i} = \begin{cases}
    \check{\boldsymbol{\mu}}  & \text{reflection} \\
    \check{\boldsymbol{\mu}}  + \frac{(\mathbf{x}^\text{BS} - \check{\boldsymbol{\mu}})^{\T} \mathbf{u}}{2 (\hat{\mathbf{p}} - \check{\boldsymbol{\mu}})^{\T} \mathbf{u} } (\hat{\mathbf{p}} - \check{\boldsymbol{\mu}}) & \text{scattering}
    \end{cases}
\end{equation}
where the time index is omitted for brevity and the auxilary variable is given by \cite{kim2020c}
\begin{equation}
    \check{\boldsymbol{\mu}} = \begin{bmatrix}
    \lbrace \mathbf{m}_{k-1} \rbrace_{1} + r^j \cos(\lbrace \mathbf{z}_{k-1}^j \rbrace_{4} + \lbrace \mathbf{m}_{k-1} \rbrace_{3}) \\
    \lbrace \mathbf{m}_{k-1} \rbrace_{2} + r^j \sin(\lbrace \mathbf{z}_{k-1}^j \rbrace_{4} + \lbrace \mathbf{m}_{k-1} \rbrace_{3}) \\
     (\lbrace \mathbf{z}_{k-1}^j \rbrace_{1} - \lbrace \mathbf{m}_{k-1} \rbrace_{4}) \sin(\lbrace \mathbf{z}_{k-1}^j \rbrace_{5})
    \end{bmatrix}
\end{equation}
and $\hat{\mathbf{p}} = [ \lbrace \mathbf{m}_{k-1} \rbrace_{1:2}^{\T} , \; 0 ]^{\T}$ denotes the estimated UE position, $\mathbf{u} = (\mathbf{x}^\text{BS} - \check{\boldsymbol{\mu}}_{k-1}) / \lVert \mathbf{x}^\text{BS} - \check{\boldsymbol{\mu}}_{k-1} \rVert$ and $r^j = (\lbrace \mathbf{z}_{k-1}^j \rbrace_{1} - \lbrace \mathbf{m}_{k-1} \rbrace_{4}) \cos(\lbrace \mathbf{z}_{k-1}^j \rbrace_{5})$. Covariance of the birth component is approximated using
\begin{align}
\mathbf{C}_{k -1}^{\text{B},i}  &= \left( [\grave{\mathbf{G}}_\mathbf{x}]^{\T} \boldsymbol{\Sigma}^{-1} \grave{\mathbf{G}}_\mathbf{x}\right)^{-1} + \acute{\mathbf{G}}_\mathbf{x}\mathbf{P}_{k-1} [\acute{\mathbf{G}}_\mathbf{x}]^{\T},
\end{align}
where the first term represents measurement uncertainty and the latter term uncertainty of the UE state estimate. In the equation, $\grave{\mathbf{G}}_\mathbf{x}$ denotes the Jacobian of $\mathbf{g}(\cdot)$ computed with respect to $\boldsymbol{\mu}_{k-1}^{\text{B},i}$ and $\acute{\mathbf{G}}_\mathbf{x}$ the Jacobian of $\boldsymbol{\mu}_{k-1}^{\text{B},i}$ computed with respect to $\mathbf{m}_{k-1}$.


\subsection{Update}

\subsubsection{Vehicle Update}\label{sec:vehicle_update}

The considered problem requires solving the data association problem since it is not known which measurement originates from which landmark. We utilize a technique based on the global nearest neighbor assignment \cite{barshalom_1995}, which provides a hard decision regarding the association of measurements to different sources. This problem can be casted as an optimal assignment problem
\begin{align}\label{eq:optimization_problem}
\text{maximize} \quad  & \text{tr} \left(\mathbf{A}^{\T} \mathbf{L} \right) \\
\text{s.t.} \quad  & \lbrace \mathbf{A} \rbrace_{i,j} \in 
\lbrace 0, \; 1 \rbrace \quad \forall \; i,j \nonumber \\ 
& \sum\nolimits_{j=1}^{n_k+m_k} \lbrace \mathbf{A} \rbrace_{i,j} = 1, \quad \forall \; i \nonumber \\ 
& \sum\nolimits_{i=1}^{n_k} \lbrace \mathbf{A} \rbrace_{i,j} \in  \lbrace 0, \; 1 \rbrace, \quad \forall \; j \nonumber 
\end{align}
where $\mathbf{A}$ and $\mathbf{L}$ are the assignment and cost matrix in respective order. The cost matrix is defined as
\begin{equation}
\mathbf{L} = \left[
\begin{matrix}
\ell^{1,1} & \ldots & \ell^{1,m_k} \\ 
\vdots &  \ddots & \vdots \\
\ell^{n_k,1}  & \ldots & \ell^{n_k,m_k}
\end{matrix}
\left|
\,
\begin{matrix}
\ell^{1,0}  & \ldots & -\infty \\ 
\vdots &  \ddots & \vdots \\
-\infty &  \ldots & \ell^{n_k,0}
\end{matrix}
\right.
\right]
\end{equation}
where the left $n_k \times m_k$ sub-matrix corresponds to detections, the right $n_k \times n_k$ diagonal sub-matrix corresponds to misdetections, and the off-diagonal elements of the right submatrix are $\log(0) = -\infty$. The optimization problem is solved using the Auction algorithm \cite{blackman_1999} and it finds the assignment matrix $\mathbf{A}^*$ that maximizes the log-likelihood. It is to be noted that based on $\mathbf{A}^*$, $J$ measurements will be associated to existing landmarks and some measurements will not be associated to any landmark. These measurements will lead to the generation of new landmarks, which was discussed in Section \ref{sec:map_prediction}.

To construct the cost matrix $\mathbf{L}$, we proceed as follows. Let us denote the combined state vector of the UE and $i$-th landmark as $\check{\mathbf{x}}^i_{k \vert k -1} = [(\mathbf{m}_{k \vert k-1})^{\T}, \; (\boldsymbol{\mu}_{k \vert k-1}^i)^{\T} ]^{\T}$. Now, the elements of the cost matrix are computed as \cite{garcia2018}:
\begin{equation}
\ell^{i,j} \mkern-4mu = \mkern-4mu \begin{cases}
    \log \left(1 - P^\text{D}(\check{\mathbf{x}}^i_{k \vert k -1})\right) & \mkern-26mu  \text{misdetection} \\
    \log \left(\frac{P^\text{D}(\check{\mathbf{x}}^i_{k \vert k -1})}{\lambda_c(\mathbf{z}^j_k)}\right) \mkern-4mu - \mkern-4mu \frac{1}{2} \mkern-2mu \left(\log((2 \pi)^V \lvert \mathbf{S}_k^i \rvert) \mkern-4mu + \mkern-4mu d \right) & \text{detection}
\end{cases}
\end{equation}
where $P^\text{D}(\check{\mathbf{x}}^i_{k \vert k -1})$ is an adaptive detection probability \cite{kim2020c}, $\lambda_c(\mathbf{z}^j_k)$ clutter intensity, $V= \mathrm{dim}(\mathbf{z}^j_k)$,  $d$ is a square of a Mahalanobis distance:\footnote{\noindent The computational complexity is reduced by applying ellipsoidal gating and $\ell^{i,j} = -\infty$ if $d \geq T_G$, where $T_G$ is the gating threshold.}
\begin{equation}
    d = \left(\mathbf{z}_k^j - \mathbf{g}(\check{\mathbf{x}}^i_{k \vert k -1})  \right)^{\T} \left(\mathbf{S}_k^i\right)^{-1} \left(\mathbf{z}_k^j - \mathbf{g}(\check{\mathbf{x}}^i_{k \vert k -1})  \right)
\end{equation}
and $\mathbf{S}_k^i$ is the innovation covariance, given by
\begin{equation}\label{eq:UE_innovation_covariance}
    \mathbf{S}_k^i = \mathbf{G}_\mathbf{x}(\check{\mathbf{x}}^i_{k \vert k -1}) \check{\mathbf{P}}_{k \vert k -1}^i \left[\mathbf{G}_\mathbf{x}(\check{\mathbf{x}}^i_{k \vert k -1})\right]^{\T} + \boldsymbol{\Sigma},
\end{equation}
in which $\check{\mathbf{P}}_{k \vert k -1}^i = \text{blkdiag}(\mathbf{P}_{k \vert k-1},\mathbf{C}_{k \vert k-1}^i)$ and $\mathbf{G_x}(\check{\mathbf{x}}_{k \vert k -1}^i)$ is the Jacobian of $\mathbf{g}(\cdot)$ with elements $\lbrace \mathbf{G_x}(\check{\mathbf{x}}_{k \vert k -1}^i) \rbrace_{i,j} = \partial g_i (\mathbf{x})/\partial {x}_j \vert_{\mathbf{x} = \check{\mathbf{x}}_{k \vert k -1}^i}$. 


 Let $l(\cdot)$ denote the optimal assignment matrix $\mathbf{A}^*$ mapping between the map object and measurement, and $J$ the number of landmarks associated to the measurements. Using $l(\cdot)$, the mean, covariance and measurements are given by
 \begin{align*}
     \breve{\mathbf{x}}_{k \vert k -1} &= \begin{bmatrix}
    {\mathbf{m}}_{k \vert k-1}^{\T},
    [{\boldsymbol{\mu}}^{l(1)}_{k \vert k-1}]^{\T}, & \cdots, & [{\boldsymbol{\mu}}^{l(J)}_{k \vert k-1} ]^{\T}
    \end{bmatrix}^{\T} \\
    \breve{\mathbf{P}}_{k \vert k-1} &= \text{blkdiag}(\mathbf{P}_{k \vert k-1},\mathbf{C}_{k \vert k-1}^{l(1)}, \cdots, \mathbf{C}_{k \vert k-1}^{l(J)}) \\
     \mathbf{g}(\breve{\mathbf{x}}_{k \vert k-1}) &= \begin{bmatrix}
    [\mathbf{g}(\check{\mathbf{x}}^{l(1)}_{k \vert k-1}) ]^{\T},  & \cdots, &  [\mathbf{g}(\check{\mathbf{x}}^{l(J)}_{k \vert k-1}) ]^{\T}
    \end{bmatrix}^{\T}, \\
    \breve{\mathbf{z}}_k &= \begin{bmatrix}
    [\mathbf{z}_k^{1} ]^{\T}, & \cdots, & [\mathbf{z}_k^{J} ]^{\T}
    \end{bmatrix}^{\T},
\end{align*}
and the first order extended Kalman filter \cite[Ch. 5.2]{sarkka_2013} can be used to update the mean and covariance 
\begin{align}
    \mathbf{S}_k &= \mathbf{G_x}(\breve{\mathbf{x}}_{k \vert k-1}) \breve{\mathbf{P}}_{k \vert k-1} \left[\mathbf{G_x}(\breve{\mathbf{x}}_{k \vert k-1})\right]^{\T} + \breve{\boldsymbol{\Sigma}}_k \\
    \mathbf{K}_k &= \breve{\mathbf{P}}_{k \vert k-1} \left[\mathbf{G_x}(\breve{\mathbf{x}}_{k \vert k-1})\right]^{\T} \mathbf{S}_k^{-1} \\
    \breve{\mathbf{x}}_{k} &= \breve{\mathbf{x}}_{k \vert k -1} + \mathbf{K}_k \left(\breve{\mathbf{z}}_k - \mathbf{g}(\breve{\mathbf{x}}_{k \vert k-1}) \right)\\
    \breve{\mathbf{P}}_{k} &= \breve{\mathbf{P}}_{k \vert k -1} - \mathbf{K}_k \mathbf{S}_k \mathbf{K}_k^{\T},
\end{align}
where $\breve{\boldsymbol{\Sigma}}_k = \mathbf{I}_{J} \otimes \boldsymbol{\Sigma}$ and $\mathbf{G_x}(\breve{\mathbf{x}}_{k \vert k -1})$ is the Jacobian of $\mathbf{g}(\cdot)$ with elements $\lbrace \mathbf{G_x}(\breve{\mathbf{x}}_{k \vert k -1}) \rbrace_{i,j} = \partial g_i (\mathbf{x})/\partial {x}_j \vert_{\mathbf{x} = \breve{\mathbf{x}}_{k \vert k -1}}$. The mean and covariance, $\mathbf{m}_{k}$ and 
$\mathbf{P}_{k}$, are obtained from $\breve{\mathbf{x}}_{k}$ and $\breve{\mathbf{P}}_{k}$ by marginalizing the other states out.

\subsubsection{Map Update}\label{sec:map_update}
Under the assumptions that the scatterers generate observations independently and that clutter and predicted RFS are Poisson, it can be shown that the posterior of the PHD can be updated using \cite{vo2006}
\begin{equation}
    D_{k}(\mathbf{x}) =  \left(1 - P_k^\text{D}(\mathbf{x})\right) D_{k\vert k-1}( \mathbf{x})+\sum_{\mathbf{z} \in \mathcal{Z}_k}D_{k}^\text{D}(\mathbf{x}; \mathbf{z}),
\end{equation}
where the first term represents objects that are undetected, the latter term is the set of detected objects and $P_k^\text{D}(\mathbf{x})$ is the short hand notation for the adaptive detection probability. Recall that $l(i)$ denotes the optimal assignment matrix $\mathbf{A}^*$ mapping between the landmark and $i$-th measurement. The GM components of measurements that are associated to a landmark are given by
\begin{align*}
    D_{k}^\text{D}(\mathbf{x}; \mathbf{z}) &=  w_{k}^{l(i)}(\mathbf{z}) \mathcal{N}\left(\mathbf{x} ; {\boldsymbol{\mu}}_{k}^{l(i)} , {\mathbf{C}}_{k}^{l(i)}\right), \\
     w_{k}^{l(i)}(\mathbf{z}) &= \frac{P_k^\text{D}(\mathbf{x}) w^{l(i)}_{k \vert k-1} \mathcal{N}(\mathbf{z}; \mathbf{g}(\check{\mathbf{x}}_{k \vert k-1}^{l(i)}), \mathbf{S}_k^{l(i)}) }{\lambda_c(\mathbf{z}) +  P_k^\text{D}(\mathbf{x}) w^{l(i)}_{k \vert k-1} \mathcal{N}(\mathbf{z}; \mathbf{g}(\check{\mathbf{x}}_{k \vert k-1}^{l(i)}), \mathbf{S}_k^{l(i)})},
\end{align*}
where ${\boldsymbol{\mu}}_{k}^{l(i)} \text{ and } {\mathbf{C}}_{k}^{l(i)}$ are obtained from $\breve{\mathbf{x}}_{k}$ and $\breve{\mathbf{P}}_{k}$ by marginalizing the other states. Accordingly, $\mathbf{g}(\check{\mathbf{x}}_{k \vert k-1}^{l(i)}) \text{ and } \mathbf{S}_k^{l(i)}$ can be obtained from $\mathbf{g}(\check{\mathbf{x}}_{k \vert k-1}) \text{ and } \mathbf{S}_k$. The number of GM components in the updated PHD is $n_{k \vert k} = n_{k \vert k - 1} + J$, where $n_{k \vert k - 1}$ is the number of GM components in the predicted PHD and $J$ the number of landmarks associated to the measurements.

\section{Numerical Evaluation}\label{sec:numerical_evaluation}

\subsection{Simulation Setup}

\begin{figure*} [t]
    \centering
  \subfloat[GOSPA \label{fig:results_a}]{%
       \includegraphics[width=5.9cm]{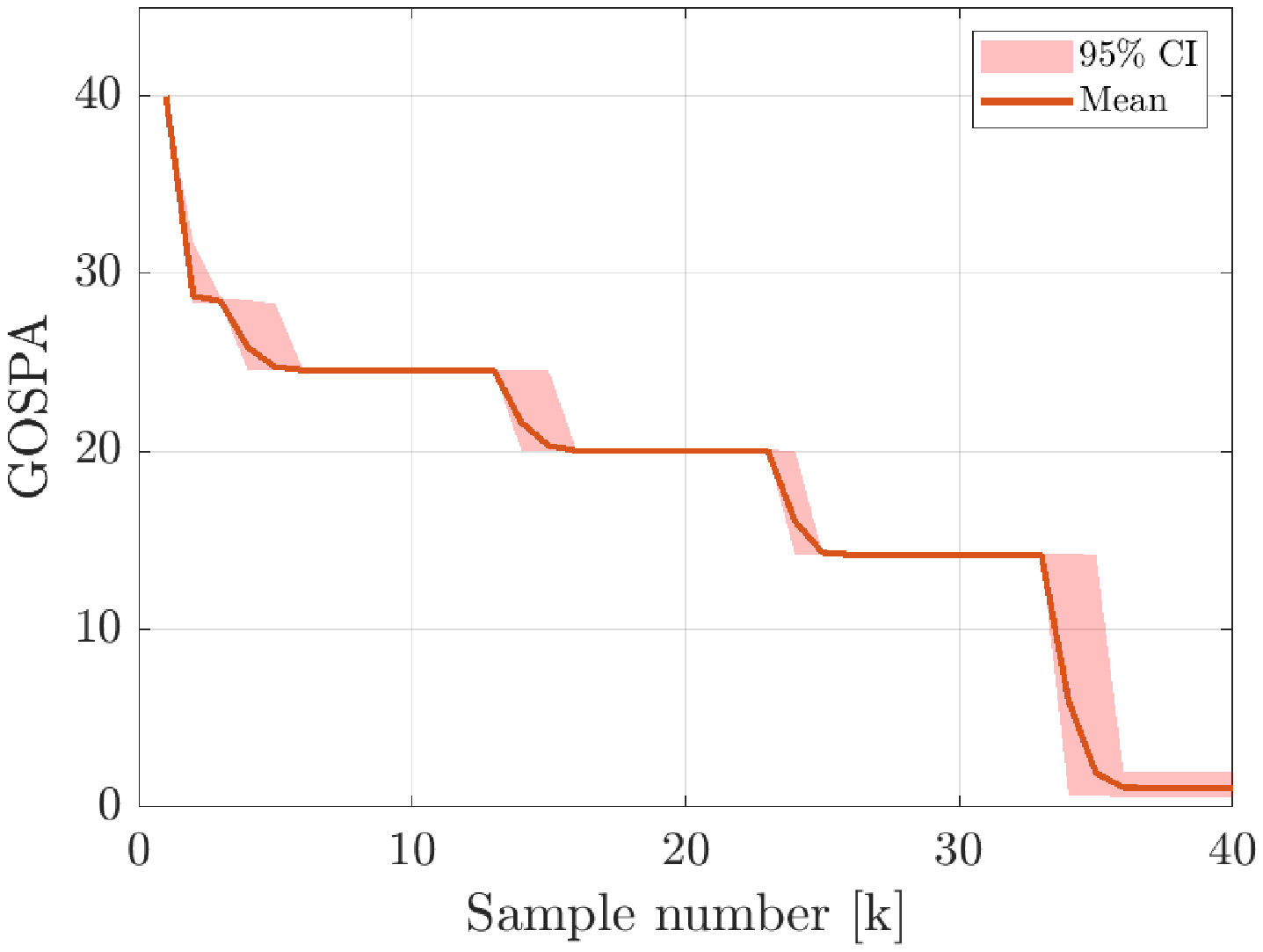}}
    \hfill
  \subfloat[Landmark localization \label{fig:results_b}]{%
        \includegraphics[width=5.9cm]{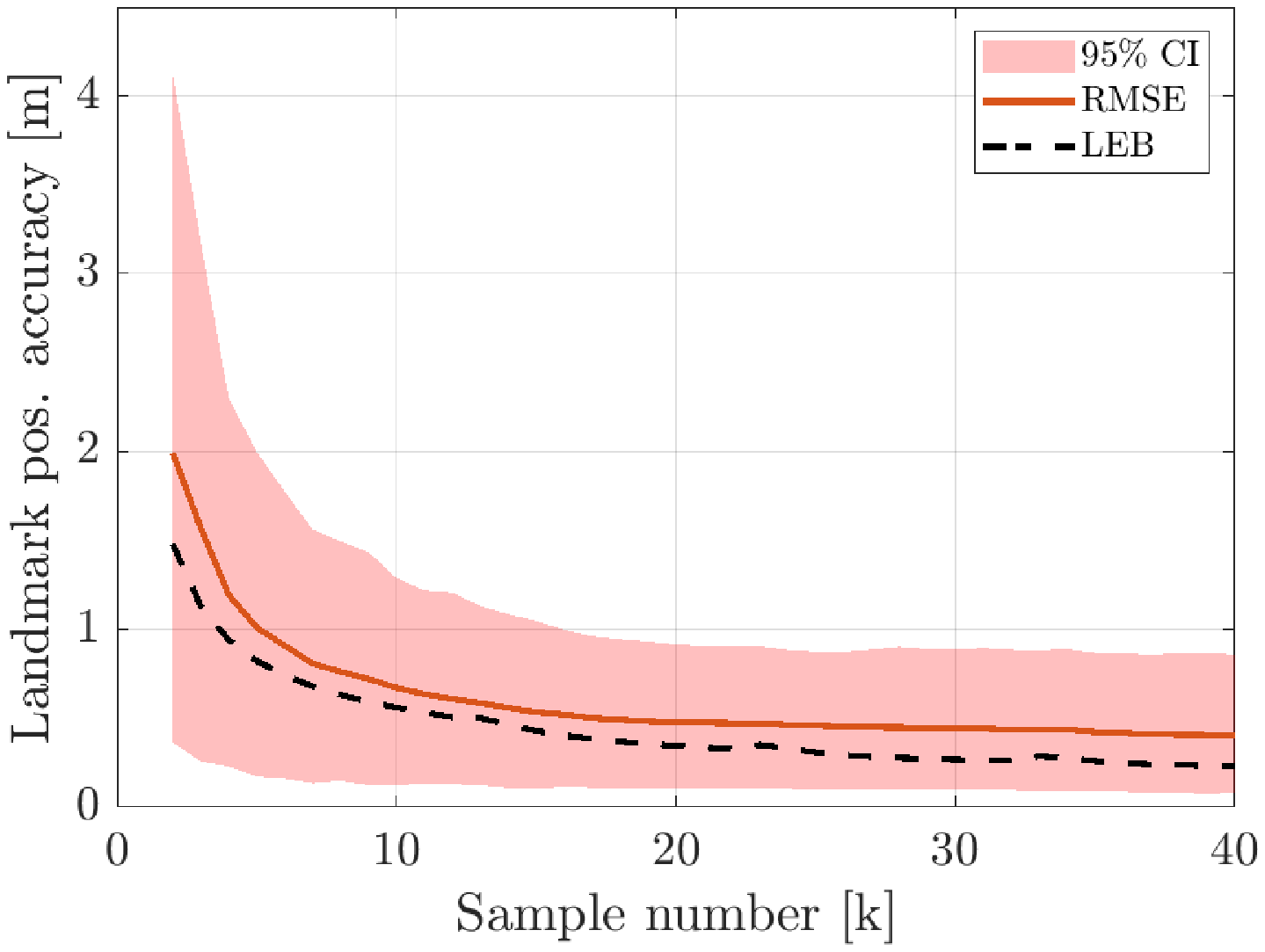}}
    \hfill
  \subfloat[Vehicle localization \label{fig:results_c}]{%
        \includegraphics[width=5.9cm]{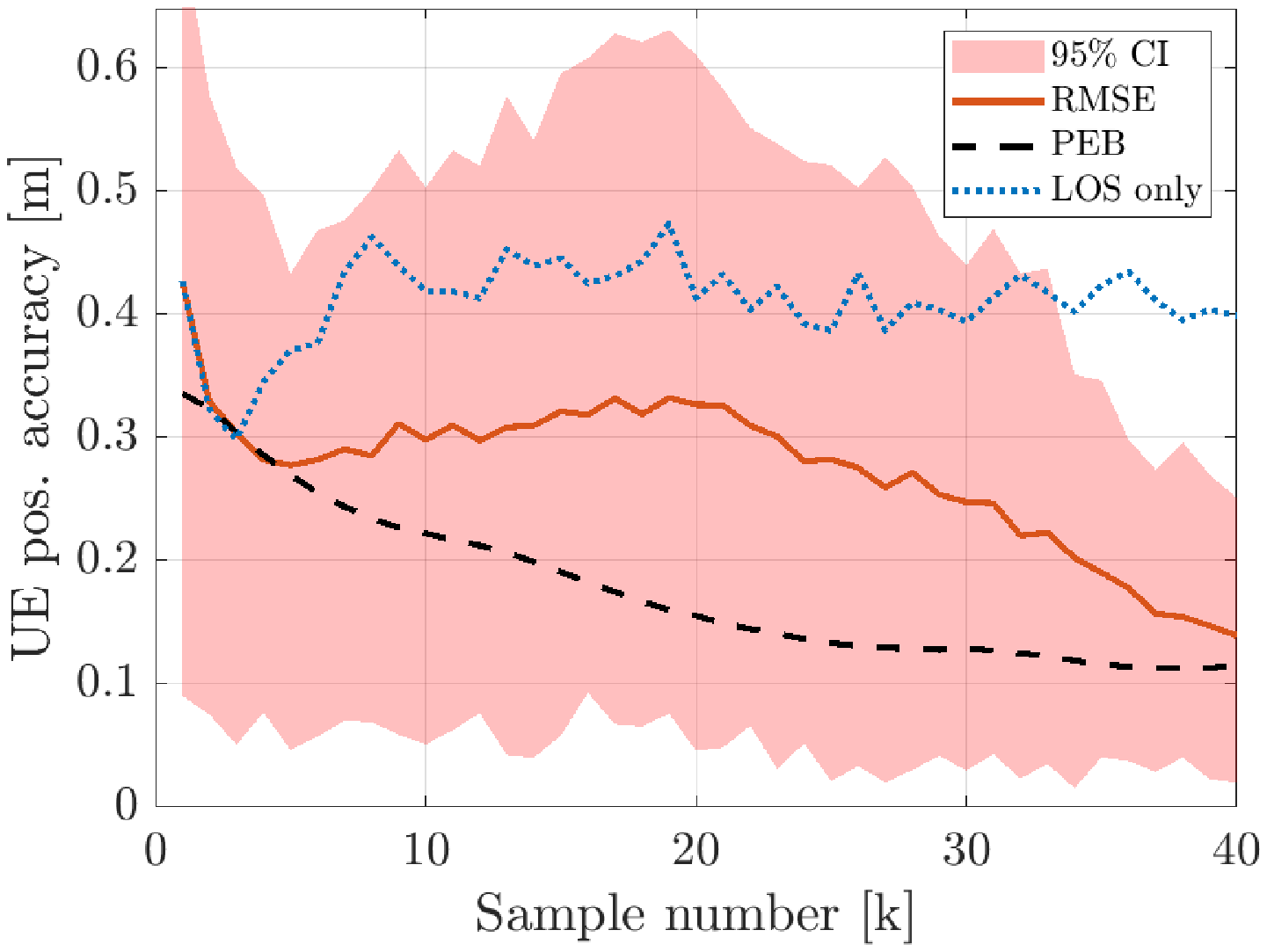}}
    \\
      \subfloat[GOSPA \label{fig:results_d}]{%
       \includegraphics[width=5.9cm]{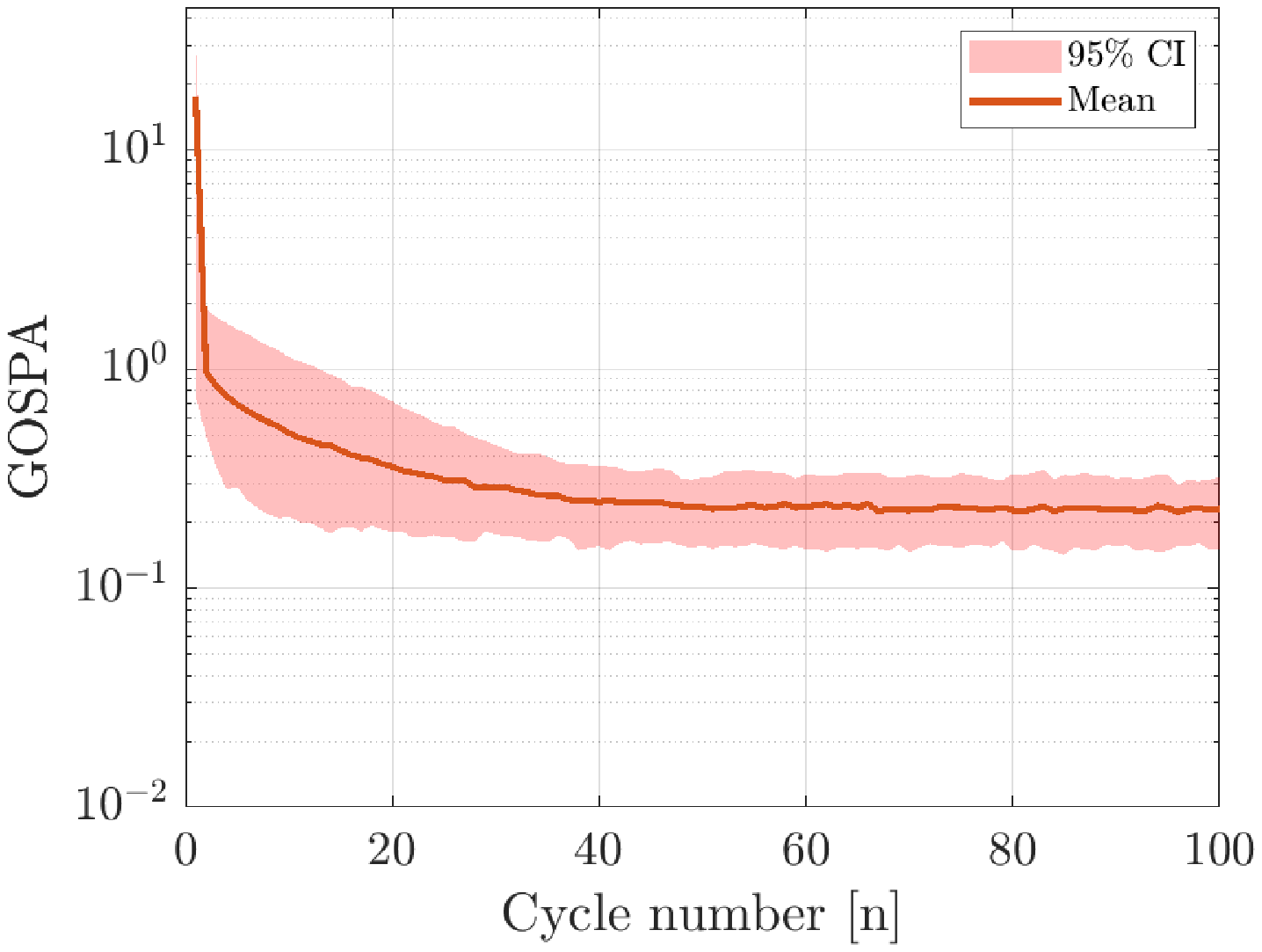}}
    \hfill
  \subfloat[Landmark localization \label{fig:results_e}]{%
        \includegraphics[width=5.9cm]{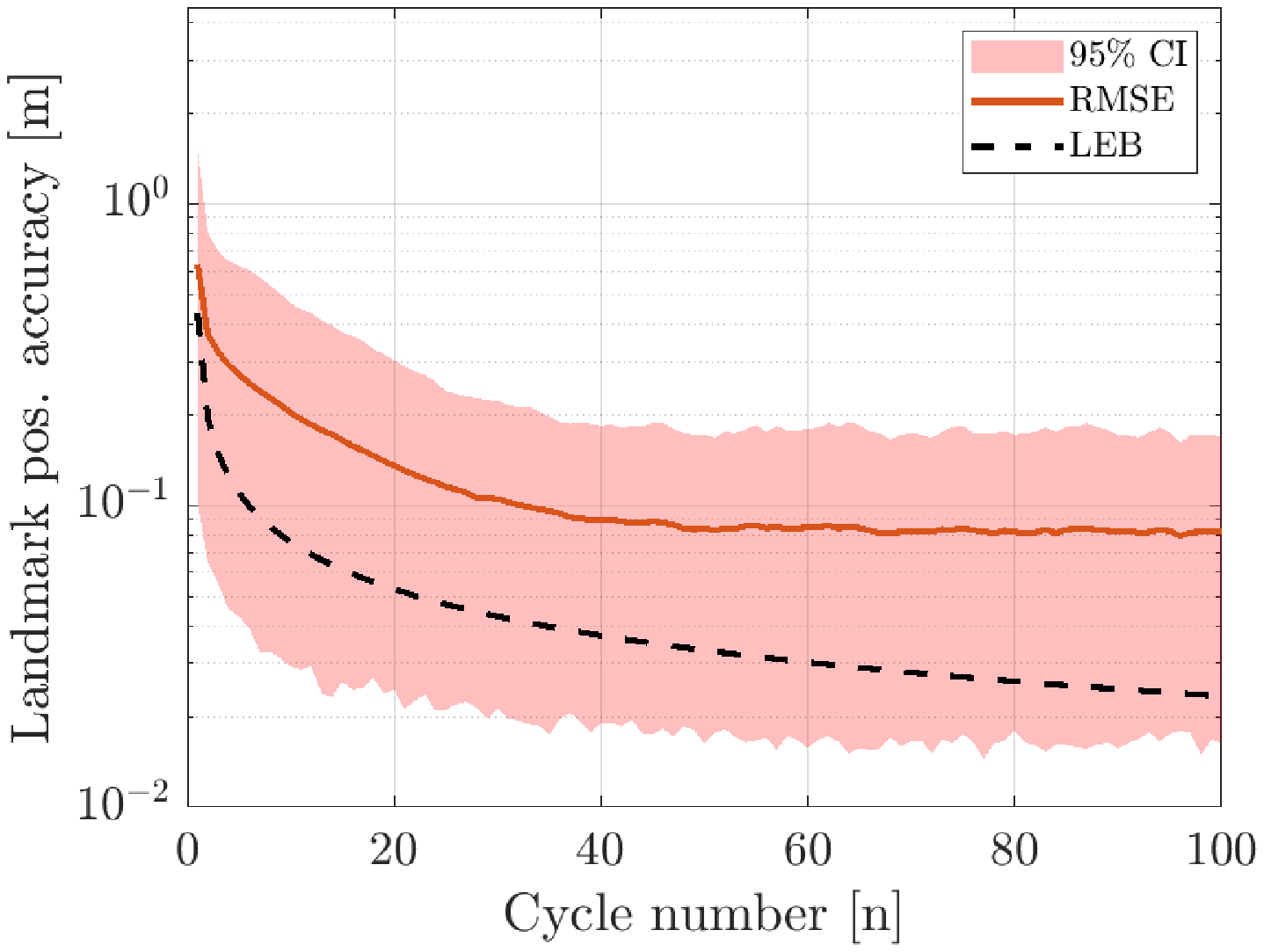}}
    \hfill
  \subfloat[Vehicle localization \label{fig:results_f}]{%
        \includegraphics[width=5.9cm]{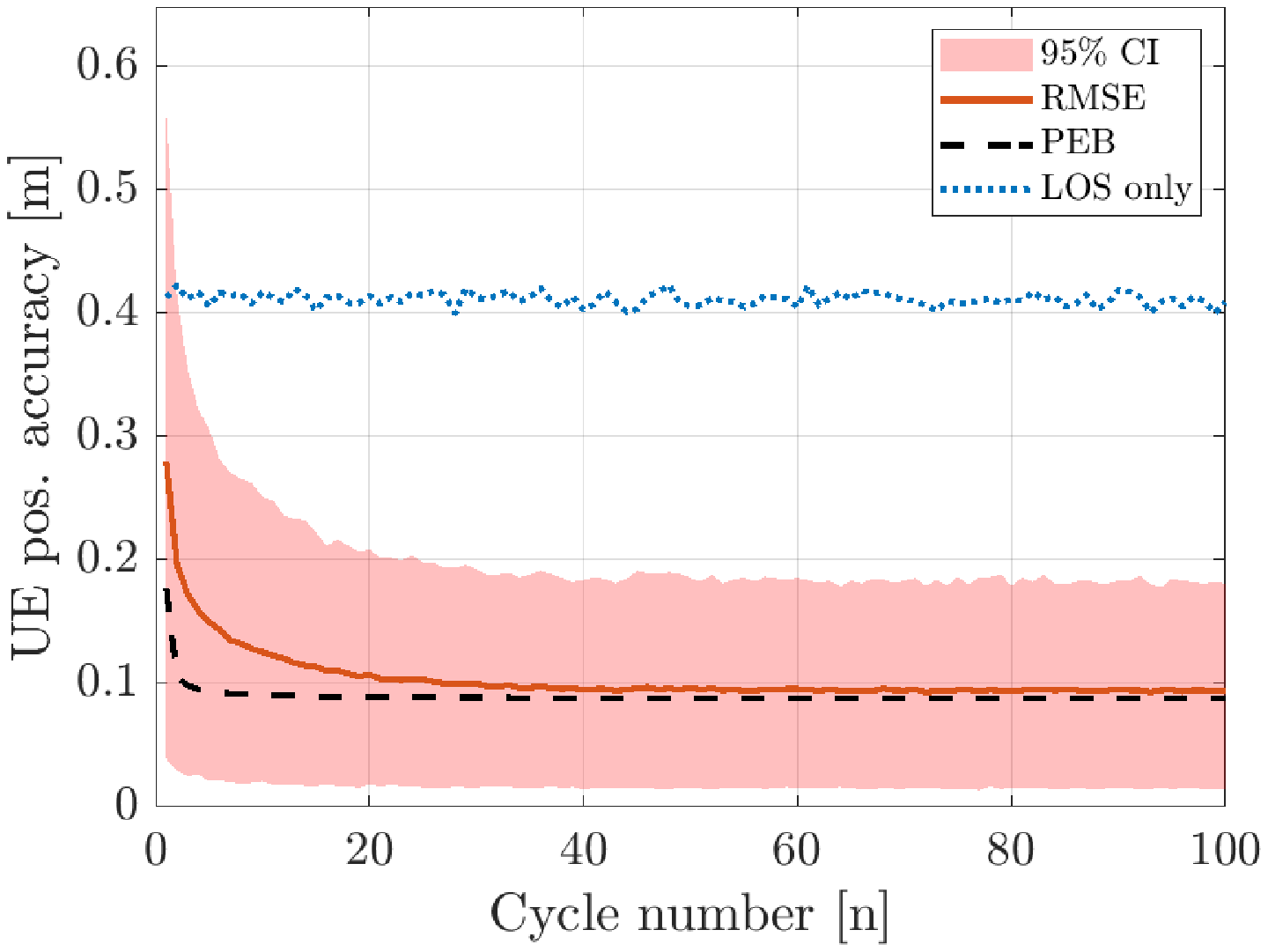}}
  \caption{The mapping and localization performance of the EK-PHD filter. On the top row, the first vehicle cycle and on the bottom, all $100$ cycles. From left to right: (a,d) GOSPA, (b,e) landmark localization, (c,f) vehicle localization. In the figures, the red line illustrates the mean and the transparent shaded area depicts the $95\%$ confidence interval of the estimates.}
  \label{fig:results}
\end{figure*}


We consider a three dimensional (3D) scenario where a single vehicle is on a circular road.  In the area, a single BS, four VAs, and four SPs are located (see Fig. 1 in \cite{kim2020c}). The vehicle state evolves according to the model in \eqref{eq:dynamic_model}, the process noise is $\mathbf{Q}^\text{UE} = \text{diag}(0.2^2 \text{ m}^2, \,0.2^2 \text{ m}^2, \,0.001^2 \text{ rad}^2, \,0.2^2 \text{ m}^2)$, the sampling interval $T=0.5 \text{ s}$ and it takes $K = 40$ samples for the vehicle to circle the road once. In total, the vehicle circles the road $100$ times. The measurements are corrupted by zero-mean Gaussian noise with covariance  $\boldsymbol{\Sigma} = \text{blkdiag}(10^{-2} \text{ m}^2, 10^{-4} \cdot \mathbf{I}_4  \text{ rad}^2)$. The vehicle state is initialized once at the beginning of the experiments using $\mathbf{m}_0 \sim \mathcal{N}(\mathbf{x}_0,\mathbf{P}_0)$, where  $\mathbf{P}_0 = \text{diag}(0.3^2 \text{ m}^2, \,0.3^2 \text{ m}^2, \,0.0052^2 \text{ rad}^2, \,0.3^2 \text{ m}^2)$ and $\mathbf{x}_0 = [v/\omega \text{ m}, \; 0 \text{ m}, \; \pi/2 \text{ rad}, \; 300 \text{ m}]^{\T}$ for which the speed and turn rate are $v = 22.22 \text{ m/s}$ and $\omega = \pi/10 \text{ rad/s}$ in respective order. The considered scenario, models and parameters, with slight variations, has been used in works that are closely related \cite{kim2020a,kim2020b,kim2020c}. 

The BS is located at $[0 \; 0 \; 40]^{\T}$ m; the VAs at: $[200 \; 0 \; 40]^{\T}$, $[0 \; 200 \; 40]^{\T}$, $[-200 \; 0 \; 40]^{\T}$, $[0 \; -200 \; 40]^{\T}$ m and; the SPs at: $[65 \; 65 \; z]^{\T}$, $[-65 \; 65 \; z]^{\T}$, $[-65 \; -65 \; z]^{\T}$, $[65 \; -65 \; z]^{\T}$ and the height of the SPs is random $z \sim \mathcal{U}(0,40)$. The SP visibility radius is $50 \text{ m}$, expectation of the Poisson distribution is $\lambda = 1$ and clutter intensity $\lambda_c(\mathbf{z}) = \lambda/(4 r \pi^4 )$ where $r=200 \text{ m}$ is the maximum sensing range. The detection, survival and birth probabilities are set to $P^\text{D} =  0.9$, $P^\text{S} =  0.99$ and $P^\text{B} =  10^{-6}$ and if the scatterer is outside the FOV, $P^\text{D} =  0$ and $P^\text{S} =  1$ to avoid problems with misdetected scatterers. Pruning, merging, capping and ellipsoidal gating are utilized to decrease the computational complexity (the used parameters are  pruning threshold is $\log(10^{-6})$, merging threshold $50$, capping number $50$ and gating threshold $T_G = 10^{-9}$). The artificial process noise of the map is set to $\tilde{\mathbf{Q}}^\text{MAP} = 10^{-4} \cdot \mathbf{I}_3 \text{ m}^2$.

The mapping accuracy is evaluated using mean of the generalized optimal subpattern assignment (GOSPA) \cite{rahmathullah_2017} and the root mean squared error (RMSE) is used to measure accuracy of the state estimates. In addition, the RMSEs are evaluated with respect to the lower bounds defined in Section \ref{sec:performance_bounds}. Overall, $100$ Monte Carlo (MC) simulations are performed and the results are obtained by averaging over the different MC simulations. 

\subsection{Simulation Results}

GOSPA, mapping and positioning accuracy for each sample of the first vehicle cycle is illustrated on the top row of Fig.~\ref{fig:results}. During the first cycle, the GOSPA progressively decreases as the reflected/scattered signals from each landmark are received enabling their localization once in the FOV. As a result, the estimated map improves gradually (see Fig.~\ref{fig:results_b}) as more measurement become available and the geometric dilution of precision decreases as the vehicle moves along the circular road. In turn, online mapping improves the position estimates considerably as illustrated in Fig.~\ref{fig:results_c}. In the figure, a filter that only relies on the LOS signal is also visualized to emphasize the benefit of Radio-SLAM. The RMSEs of the heading and clock bias estimates show similar trends but the results are omitted for brevity. 

In the following cycles, the vehicle has already estimated the map and the additive information can be used to enhance the estimates which in turn improves vehicle tracking. GOSPA, mapping and positioning accuracy of the vehicle are illustrated on the bottom row of Fig.~\ref{fig:results} as a function of cycle number. The results imply that the map converges around $50$ cycles after which the estimates do not improve anymore since covariance of the map is inflated by artificial noise as discussed in Section \ref{sec:map_prediction}. Despite the fact that the filter does not converge as fast as predicted by the LEB and it does not achieve the bound, the system performance is still better with the artificial noise than without it. In addition, the artificial noise has a neglectable impact to the UE state estimates  since the filter nearly achieves the PEB as shown in Fig.~\ref{fig:results_f}.

\begin{figure}[tb]
  \centering
  \centerline{\includegraphics[width=8.5cm]{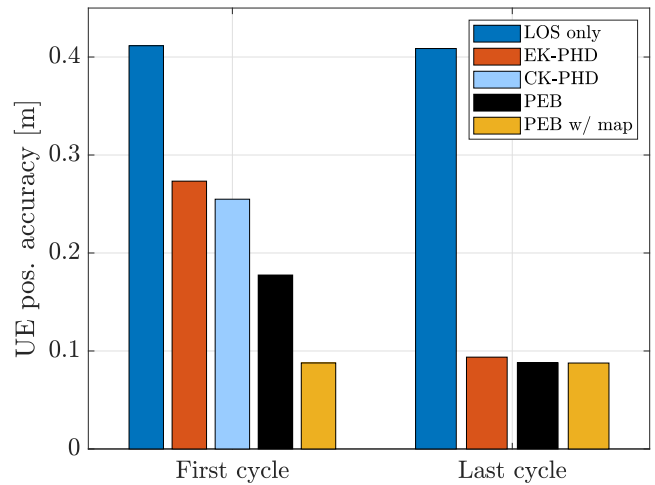}}
\caption{
RMSE and PEB during the first and last cycle of the simulations. 
}
\label{fig:result_summary}
\end{figure}

Next, performance of proposed filter is compared to the CK-PHD \cite{kim2020b}. Fig.~\ref{fig:result_summary} illustrates the RMSE of the UE position without mapping the environment (LOS only), using the method proposed in this paper (EK-PHD) and comparing it to the CK-PHD. Clearly, mapping the environment and using the multipath signals improves the positioning accuracy of both filters. The CK-PHD has a slight advantage over EK-PHD but the reason is not how the Gaussian approximation is formed. Instead, the mean and covariance estimates of the birth components are better with the CK-PHD because it uses a nonlinear optimization method. This results to improved tracking performance but the gains diminish as time evolves and the significance of the initial estimate vanishes. It is to be noted that if tracking accuracy would be the most important evaluation criteria, we could easily utilize the same nonlinear optimization method. As a result, the EK-PHD accuracy would improve and the two methods would yield comparative performance already from the very beginning. 

The results demonstrate that in theory, the map can be estimated so accurately that the UE tracking is as accurate, as with a system that has perfect knowledge of the map (see PEB and PEB w/ map in Fig.~\ref{fig:result_summary}). Furthermore, it is demonstrated that the EK-PHD can nearly achieve this bound as long as a sufficient number of measurement are available. In real-world scenarios, it is unrealistic that a single vehicle would circle around a specific city block tens of times to build up the map. However, the numerous vehicles on the road can collectively gather measurements and build up the map in a collaborative manner \cite{kim2020b,kim2020c}. 

\begin{table}[]
    \centering
    \caption{Computation time in milliseconds of the prediction and update steps of the filters. The results are obtained using a Matlab implementation running on a MacBook Pro with a 2.6 GHz 6-Core Intel Core i7 processor and 16 GB of memory.}
        \begin{tabular}{ |c|c|c|c| } 
         \hline
         Filter & Prediction & Update & Total \\ \hline 
         EK-PHD (proposed) & $0.4$ & $2.7$ & $3.1$ \\ 
         CK-PHD \cite{kim2020b} & $420.7$ & $141.1$ & $561.8$ \\ 
         \hline
        \end{tabular}
        
    \label{tab:cpu_time}
\end{table}

The most significant benefit of the proposed system is the low computational overhead of the filter. It is not straightforward to compare different implementations to one another, but if we directly measure the runtime of the EK-PHD and compare it to the CK-PHD \cite{kim2020b}, we are able to clock a $181$  times improvement with respect to CK-PHD as tabulated in Table~\ref{tab:cpu_time}. The most notable difference between the algorithms is how the Gaussian approximation is formed: the EKF relies on local linearization, whereas the CKF uses sigma points that are propagated through the nonlinearity. Another notable difference is the used birth process, the CKF utilizes a nonlinear optimization algorithm \cite{kim2020c} that relies on numerical differentiation to approximate the required Jacobians. The method proposed in this paper directly estimates the mean from the measurements and the Jacobians are computed in closed form. 






\section{Conclusions}
 
 We have proposed a low complexity EK-PHD filter for Radio-SLAM and demonstrated that the filter nearly achieves the PEB when a sufficient number of measurements are available for mapping the environment. The high tracking performance in combination with low computational overhead are attractive features, making the EK-PHD filter a viable option for performing real-time Radio-SLAM in future 5G and beyond networks.
 
\section*{Acknowledgement}

This work is partially funded by the Academy  of Finland Grants \#319994, \#323244, \#328214 and \#338224. This material is also based upon work supported by the Vinnova 5GPOS project under grant 2019-03085, by the Swedish Research Council under grant No. 2018-03701, and by the Wallenberg AI, Autonomous Systems and Software Program (WASP) funded by Knut and Alice Wallenberg Foundation.

\bibliographystyle{IEEEtran}
\bibliography{Bibliography}

\end{document}